\documentstyle[twoside,fleqn,tm_espcrc2]{article}

\newcommand{\AmS}{{\protect\the\textfont2
  A\kern-.1667em\lower.5ex\hbox{M}\kern-.125emS}}

\title{\vspace{-1cm}
       \hspace{13.2cm} {\small Bicocca-FT-00-20} \\
       \hspace{13cm}  \\
Non--perturbative scaling tests of twisted mass QCD
\thanks{Based on a poster by M.~Della~Morte presented at the International
Symposium on Lattice Field Theory, August 2000, Bangalore, India.}}

\author{M. Della Morte\address{Dipartimento 
                           di Fisica,
                           Universit\`a di Milano--Bicocca and INFN
                           Sezione di Milano, Milano, Italy},
        R. Frezzotti$^{\rm a}$,  
        J. Heitger\address{Westf\"alische Wilhelms-Universit\"at M\"unster,
        Institut f\"ur Theoretische Physik, M\"unster, Germany} and 
        S. Sint \address{CERN, Theory Division, Geneva, Switzerland}}

\def\defi             {\stackrel{\rm def}{=}}

\newcommand\be{\begin{equation}}
\newcommand\ee{\end{equation}}
\newcommand\bea{\begin{eqnarray}}
\newcommand\eea{\end{eqnarray}}
\newcommand{\psibar}{\bar{\psi}}

\newcommand{\rmR}{{\rm R}}
\newcommand{\muq}{\mu_{\rm q}}
\newcommand{\mq}{m_{\rm q}}

\begin{document}

\begin{abstract}
 We present a scaling study of lattice QCD with O($a$) improved Wilson 
 fermions and a chirally twisted mass term. In order to get precise results 
 with a moderate computational effort, we have considered a
 system of physical size of $0.75^3 \times 1.5~{\rm fm}^4$ with Schr\"odinger 
functional boundary conditions in the quenched approximation.
Looking at meson observables in the pseudoscalar and vector channels, 
we find that O($a$) improvement is effective and residual cutoff effects are 
fairly small.  
\end{abstract}

\maketitle

\setcounter{footnote}{0}
\section{Introduction}
Twisted mass QCD (tmQCD) was introduced in~\cite{FGSW} to solve the
problem of unphysical fermion zero modes in lattice QCD with two degenerate 
flavours of Wilson fermions. The fermion 
lattice tmQCD action reads:
\be \label{S_latt}
S_F = 
a^4 \sum_x \overline{\psi}(x) \left( D +m_0 + i\muq \gamma_5 \tau^3\right) \psi(x) \; ,
\ee
where $D$ is the O($a$) improved Wilson lattice 
regularization\footnote{See ref.~\cite{O(a)} for notation and details.} of 
$D\!\!\!\!/$
and $m_0 \equiv 1/2\kappa - 4$ and $\muq$ are two real quark mass parameters.
The lattice symmetries and power counting determine the
counterterm structure for renormalization and O($a$) improvement of the infinite
volume theory. The physical equivalence of tmQCD to QCD in the continuum limit 
has been argued in~\cite{FGSW} and will be further detailed in~\cite{paper1}.

We study tmQCD in a four dimensional box with Schr\"odinger functional
(SF) boundary conditions. We wish to check  
that after on-shell O($a$) improvement \`a la Symanzik~\cite{Sym}
a few renormalized quantities related to meson physics approach the
continuum limit with residual scaling violations that are small and compatible
with being O($a^2$). 
Following ref.~\cite{Jo}, we keep the spatial size 
of the box moderately small
($L \simeq 0.75$~fm) and choose $T=2L$.

The transfer matrix of lattice tmQCD with action~(\ref{S_latt}) and $c_{\rm SW}=0$ can be constructed
in close analogy to ref.~\cite{Lu77} and turns out to be self-adjoint and strictly positive
for $|\kappa|<1/6$~\cite{paper1}. Following refs.~\cite{SchSte,SFYM}, the Schr\"odinger
functional in tmQCD can be conveniently defined as the integral kernel of an integer power
$T/a$ of the transfer matrix~\cite{paper1}. It has an Euclidean representation given by
\be \label{SF_tmQCD}
{\cal Z} [ \rho',\overline{\rho}',C';\rho,\overline{\rho},C]\; = \;
\int_{\rm fields}  e^{-S[U,\psi,\psibar]} \; 
\ee
and can hence be considered as a functional of the fields at the Euclidean times $0$ and $T$.
The structure of the transfer matrix implies that the boundary conditions for gauge and quark 
fields are the same  as in the standard framework ($\muq=0$).
The action $S[U,\psibar,\psi]$ in eq.~(\ref{SF_tmQCD}) is the sum of the usual plaquette pure gauge action
and the quark action, which takes the same form as in eq.~(\ref{S_latt}), provided one adopts
the notational conventions of Subsect. 4.2 of~\cite{O(a)}.


The tmQCD Schr\"odinger functional is expected to be finite after the standard couplings and boundary
renormalization~\cite{paper1}. 
At order $a$ new boundary counterterms proportional to $a\muq$ arise~\cite{paper1}.
Their effect beyond tree--level will be neglected in this study.

\section{Correlation functions}

In order to motivate our choice of observables, we recall some properties
of  renormalized tmQCD. For simplicity, we first consider the theory 
with boundary conditions that are periodic in space and as specified in
\cite{Lu77} in time direction.
The renormalization scheme (R) of tmQCD can be chosen~\cite{paper1} to be consistent with the 
Ward identities of the flavour chiral symmetries:
\begin{eqnarray} \label{ren_WI}
\partial_{\mu} (A_\rmR)_{\mu}^a &=& 2m_\rmR (P_\rmR)^a + 
\delta^{a3}i\mu_\rmR (S_\rmR)^0 \label{tm_PCAC} \\
\partial_{\mu} (V_\rmR)_{\mu}^a &=&-2 \mu_\rmR \varepsilon^{ab3} (P_\rmR)^b \label{tm_PCVC}
\end{eqnarray}  
where the above relations are to be understood as operator insertions\footnote{The quark bilinears are
defined in the standard way, with flavour matrices $\tau^a/2$ (for $a=1,2,3$) or $\tau^0=1$.} 
in the correlation functions and $\varepsilon$ is fully antisymmetric with $\varepsilon^{123}=1$. Note that
our chirally twisted parameterization of QCD involves two renormalized mass parameters, $m_\rmR$ and $\mu_\rmR$.

It can be shown \cite{FGSW,paper1} that, within a certain subset of the renormalization schemes
that preserve the above Ward identities, the {\em on-shell} correlation functions of tmQCD can be
mapped onto the ones of standard QCD in a related renormalization scheme. In the latter scheme 
the flavour chiral Ward identities (at the operator level) read:
\be
\partial_{\mu} (A_\rmR)^a_{\mu}=2m'_\rmR (P_\rmR)^a \; , 
\quad\quad \partial_{\mu} (V_\rmR)_{\mu}^a=0 
\ee
for $a=1,2,3$ with the mass $m'_\rmR$ satisfying:
\be \label{alpha_def}
(m'_\rmR)^2 = m_\rmR^2 + \mu_\rmR^2 \; , 
\quad \tan (\alpha) = \mu_\rmR/m_\rmR \; . 
\ee
Here $\alpha$ is an unphysical angle which just specifies the quark mass parameterization.  
An example of this mapping is given by:
\bea \label{map_ex}
\langle (A_\rmR)^2_0(x) (P_\rmR)^2(y) \rangle_{\rm c}\{g_\rmR,m_\rmR',0\} = \nonumber \\
\langle (A'_\rmR)^2_0(x) (P'_\rmR)^2(y) \rangle_{\rm c}\{g_\rmR,m_\rmR,\mu_\rmR\} \; ,
\eea
where $x \neq y$ and by definition:
\bea \label{primed1}
(A'_\rmR)_{\mu}^a &=& \cos(\alpha) (A_\rmR)_{\mu}^a + \sin(\alpha)\varepsilon^{ab3} (V_\rmR)_{\mu}^b \nonumber \; , \\
(P'_\rmR)^a &=& (P_\rmR)^a  \quad\quad\quad a=1,2 \; .
\eea
It is understood that in eq.~(\ref{map_ex}) the l.h.s. must be evaluated
at renormalized couplings $\{g_\rmR,m_\rmR',0\}$
and the r.h.s at renormalized couplings $\{g_\rmR,m_\rmR,\mu_\rmR\}$, with the quark mass
couplings related via eq.~(\ref{alpha_def}). Note that the above mapping takes the same form
as at the classical level, where it is obtained via~\cite{FGSW}
$$
\psi'_{\rm cl} = e^{ i \alpha \gamma_5 \tau^3 /2 } \psi_{\rm cl} \; , \quad\quad
\psibar'_{\rm cl} = \psibar_{\rm cl} e^{ i \alpha \gamma_5 \tau^3 /2 } \; .
$$
In the following we will also need to consider, again for $a=1,2$ only, the primed fields:
\bea \label{primed2}
(V'_\rmR)_{\mu}^a \; &=& \cos(\alpha) (V_\rmR)_{\mu}^a + \sin(\alpha)\varepsilon^{ab3} (A_\rmR)_{\mu}^b \nonumber \; , \\
(T'_\rmR)_{\mu\nu}^a &=& (T_\rmR)_{\mu\nu}^a  \quad\quad\quad a=1,2\; . 
\eea

Coming back to our scaling tests of tmQCD with SF boundary conditions, we focus on a few 
renormalized quantities that can be extracted from the following correlators:
\bea \label{SF_main_corr1}
f_{\rm{R,A'}}^{22}(x_0)&=&- \langle( A'_\rmR)^2_0(x){\cal O}_5^2 \rangle /\sqrt{f_1^{22}} \nonumber \; ,\\
f_{\rm{R,P'}}^{22}(x_0)&=&- \langle( P'_\rmR)^2(x){\cal O}_5^2 \rangle /\sqrt{f_1^{22}} \; 
\eea
and (with sum over $k=1,2,3$ understood)
\bea \label{SF_main_corr2}
k_{\rm{R,V'}}^{22}(x_0)&=& -\frac{1}{3}
 \langle( V'_\rmR)_k^2(x){\cal Q}_k^2 \rangle /\sqrt{f_1^{22}} \nonumber \; , \\
k_{\rm{R,T'}}^{22}(x_0)&=& -\frac{1}{3}
\langle( T'_\rmR)_{k0}^2(x){\cal Q}_k^2 \rangle /\sqrt{f_1^{22}} \; ,
\eea 
where ${\cal O}_5^a$ and ${\cal Q}_k^a$ are the SF boundary fields 
defined in ref.~\cite{Jo} at zero Euclidean time\footnote{The analogous fields
localized on the other SF time-boundary are labelled with a prime.}. 
In the above correlators the division by the square root of the quantity 
\be \label{f1_def}
f_1^{22}= -\frac{1}{L^6}\langle{\cal O'}_5^2 {\cal O}_5^2\rangle \; ,
\ee
takes care of the boundary field renormalization. 

In the quantum mechanical representation of the renormalized correlation functions 
that are obtained from eq.~(\ref{SF_main_corr1}) and eq.~(\ref{SF_main_corr2}), the
only non--vanishing contributions (up to cutoff effects) arise when inserting
states with vacuum quantum numbers at Euclidean times between $x_0$ and $T$, and
states with pion and $\rho$--meson quantum numbers, respectively, at Euclidean times
between $0$ and $x_0$. Let us consider for instance the SF correlator
$f_{\rm R,A'}^{22}$:
the matrix elements between intermediate states $| k \rangle $ and $| n \rangle $
\be \label{matr_el}
\langle k | \cos(\alpha) (A_\rmR)_0^2 
- \sin(\alpha) (V_\rmR)_0^1 | n \rangle \; ,
\ee
which enter its quantum mechanical representation, are independent of the SF
boundary conditions and also appear in the quantum mechanical representation
of the r.h.s. of eq.~(\ref{map_ex}). The equality of the correlators in eq.~(\ref{map_ex})
implies the equality of their quantum mechanical representations for arbitrary values
of $(x_0-y_0)$. From this equality it follows that in tmQCD  the 
matrix elements of the form (\ref{matr_el}) with the states $| k \rangle $ and
$| n \rangle $ carrying vacuum and pion quantum numbers, respectively, are non-zero.
Analogous arguments hold for the remaining SF correlators 
in eqs.~(\ref{SF_main_corr1})--(\ref{SF_main_corr2}). The fields
${\cal O}_5^a$ and ${\cal Q}_k^a$ are chosen so that the corresponding
SF boundary states have a non--zero overlap with the pion and the $\rho$--meson states. 


\section{Definition of the observables}
In this study we consider as observables some ratios of the previously introduced SF 
finite volume correlators that are expected to
approach a well defined {\em continuum limit}\footnote{
The symbol $\tilde{\partial}_{\mu}$ denotes the symmetric lattice derivative.
}:
\bea 
m_{\rm{PS}}&=& \label{obs1} 
 \frac{\tilde{\partial}_0 f_{\rm R,P'}^{22}(x_0)}{f_{\rm R,P'}^{22}(x_0)} \; , \quad {x_0=T/2} \; , \\
\widetilde{m}_{\rm {PS}}&=& 
 \frac{\tilde{\partial}_0 f_{\rm R,A'}^{22}(x_0)}{f_{\rm R,A'}^{22}(x_0)}\; , \quad {x_0=T/2} \; , \\
m_{\rm{V}}&=& 
 \frac{\tilde{\partial}_0 k_{\rm R,V'}^{22}(x_0)}{k_{\rm R,V'}^{22}(x_0)} \; , \quad {x_0=T/2} \; ,  \\ 
\widetilde{m}_{\rm{V}}&=& 
 \frac{\tilde{\partial}_0 k_{\rm R,T'}^{22}(x_0)}{k_{\rm R,T'}^{22}(x_0)} \; , \quad {x_0=T/2} \; ,  \\
\eta_{\rm{PS}}&=& 
 C_{\rm{PS}} f_{\rm R,A'}^{22}(x_0)\; , \quad {x_0=T/2} \; , \\
\widetilde{\eta}_{\rm{PS}}&=& 
 \widetilde{C}_{\rm{PS}} f_{\rm R,A'}^{22}(x_0)\; , \quad {x_0=T/2} \; , \\
\eta_{\rm{V}}&=& 
 C_{\rm{V}} k_{\rm R,V'}^{22}(x_0)\; , \quad {x_0=T/2} \; , \\
\widetilde{\eta}_{\rm{V}}&=& \label{obs8} 
 \widetilde{C}_{\rm{V}} k_{\rm R,V'}^{22}(x_0)\; , \quad {x_0=T/2} \; .
\eea
%
As $T = 2L \to \infty$, $m_{\rm{PS}}$ and $m_{\rm{V}}$ yield estimators of the pion and $\rho$-meson
mass, respectively. The constants $C_{\rm{PS}}$ and $C_{\rm{V}}$ are defined as in \cite{Jo} in terms
of $m_{\rm{PS}}$ and $m_{\rm{V}}$. $C_{\rm{PS}}$ is such that $\eta_{\rm{PS}} \to F_\pi$ 
as $T = 2L \to \infty$. The quantity $\eta_{\rm{V}}$ is not related
to the decay of the $\rho$--meson because of its unphysical normalization. Analogous considerations
hold for the alternative scaling quantities labelled with a "tilde".

The $f_{\rm R}$- and $k_{\rm R}$ correlators entering our observables are defined in 
eqs.~(\ref{SF_main_corr1})--(\ref{f1_def}) via eqs.~(\ref{primed1})--(\ref{primed2}) and
\bea \label{ren_bilin}
(A_{\rmR})^a_{\mu} &=& Z_{\rm{A}}(1+b_{\rm{A}}a\mq)(A_{\rm{I}})^a_{\mu} \nonumber \; ,\\
(V_{\rmR})^a_{\mu} &=& Z_{\rm{V}}(1+b_{\rm{V}}a\mq)(V_{\rm{I}})^a_{\mu} \nonumber \; ,\\
(P_{\rmR})^a &=& Z_{\rm{P}}(1+b_{\rm{P}}a\mq)(P_{\rm{I}})^a  \nonumber \; , \\
(T_{\rmR})^a_{\mu\nu} &=& Z_{\rm{T}} (1+b_{\rm{T}} a \mq)(T_{\rm{I}})_{\mu\nu}^a \; ,
\eea
where (restricting attention to $a=1,2$)
\bea \label{impr_bilin}
(A_{\rm{I}})^a_{\mu}&=&
 A_{\mu}^a+c_{\rm{A}} a \tilde{\partial}_{\mu} P^a+
 a\muq \tilde{b}_{\rm{A}} \varepsilon^{ab3} V_{\mu}^b \nonumber \; ,\\
(V_{\rm{I}})^a_{\mu}&=&
 V_{\mu}^a+c_{\rm{V}} a \tilde{\partial}_{\nu} T^a_{\mu\nu}+
 a\muq \tilde{b}_{\rm{V}} \varepsilon^{ab3} A_{\mu}^b \nonumber \; ,\\
(P_{\rm{I}})^a&=&P^a \nonumber \; , \\
(T_{\rm{I}})_{\mu\nu}^a&=&T_{\mu\nu}^a + c_{\rm{T}}a(
 \tilde{\partial}_{\mu}V_{\nu}^a -\tilde{\partial}_{\nu}V_{\mu}^a )\; .
\eea
For the definition of the bare lattice fields $A_{\mu}^a$, $P^a$, $V_{\mu}^a$ and $T_{\mu\nu}^a$
we follow Sect.~2 of ref.~\cite{O(a)pert}. The suffix $\rm{I}$ refers to the O($a$) improvement
of these fields, which together with O($a$) improvement of the action, eq.~(\ref{S_latt}), implies
that -in the limit $T \to \infty$- the observables in eqs.~(\ref{obs1})--(\ref{obs8}) deviate from
their continuum limit by O($a^2$) cutoff effects. The $\tilde{b}$ coefficients multiply counterterms
that are needed to subtract (bulk) cutoff effects of order $a\muq$.

 \section{Results}

We choose non--perturbative {\em mass--independent} renormalization conditions for the couplings and the
observables of interest. As $L/a$ is increased, the relation between $\beta=6/g_0^2$ and $a/r_0$ \cite{r0scale}
is employed in order to keep the physical size of the box fixed at $L=1.49 r_0$, with $r_0$ being a 
hadron scale of order $0.5$~fm. The mass parameter $\kappa$ is tuned so to keep fixed the PCAC
renormalized mass in units of $L$:
\be \label{m_ren}
Lm_{\rmR} \defi 
 \frac{L}{a} \left\{ \frac{Z_{\rm{A}}}{Z_{\rm{P}}}
\frac{\tilde{\partial}_0f_{\rm{A}}^{22}}{2f_{\rm{P}}^{22}} \right\}
= 0.020 \; ,
\ee
where:
\bea
Z_{\rm A} f_{\rm A}^{22} &=& - \langle (A_{\rmR})_0^2(x) {\cal{O}}_5^2 \rangle \; , \nonumber \\
Z_{\rm P} f_{\rm P}^{22} &=& - \langle (P_{\rmR})^2(x) {\cal{O}}_5^2 \rangle \; .
\eea
The mass parameter $\muq$ is chosen so to fulfill the condition: 
\be \label{mu_ren}
L\mu_{\rmR} \defi Z_{\rm{P}}^{-1} L\muq = 0.153 \; .
\ee 
This choice is justified by the exact {\em lattice} Ward identity
${\partial}^{\ast}_{\mu} \widetilde{V}_{\mu}^2=2 \muq P^1$, where
$\widetilde{V}_{\mu}^a$ is the one--point split vector current that
is conserved at $\muq =0$ and ${\partial}^{\ast}_\mu$ denotes the backward
lattice derivative.
The angle $\alpha$, given by $\tan(\alpha) = \mu_{\rmR}/m_{\rmR}$,
is close to $\pi / 2$. 

\small
\begin{table}[htb]
\vspace{-0.3cm}
\begin{tabular}{ccccc}
\hline
  $\beta$ &  $\muq$ & $\kappa$ & $L/r_0$   & $Lm_{\rmR}$\\
\hline
   6.0    &  0.01    & 0.134952 &  1.490(6) & 0.0228(23)\\
   6.14   &  0.00794 & 0.135614 &  1.486(7) & 0.0203(30)\\
   6.26   &  0.00659 & 0.135742 &  1.495(7) & 0.0201(23)\\
   6.47   &  0.00493 & 0.135611 &  1.488(7) & 0.0180(24)\\
\hline
\end{tabular}
\caption{{\em The bare and renormalized parameters in our simulations. 
As for $L/a$ and $L\mu_{\rmR}$ see the text.}}
\label{tab:sim1}
\vspace{-0.5cm}
\end{table}
\normalsize

The renormalization factors $Z_{\rm{A}}$, $Z_{\rm{V}}$ and $Z_{\rm{P}}$ were
determined in~\cite{O(a)2} and~\cite{ZP}, where $Z_{\rm{P}}$ is
given as a function of $\beta$ at the scale $L_0=1.436 r_0$.
As for the improvement coefficients, we employ non--perturbative estimates of $c_{\rm{SW}}$,
$c_{\rm{A}}$ \cite{NP_O(a)} and $b_{\rm{V}}$ \cite{O(a)pert}. Moreover, we use 1--loop estimates
of $c_{\rm{V}}$, $c_{\rm{T}}$, $b_{\rm{A}}$, $b_{\rm{P}}$, $b_{\rm{T}}$ \cite{O(a)pert} as well as
of the SF-boundary coefficients $c_{\rm t}$ and $\tilde{c}_{\rm t}$ \cite{SFYM}. The coefficients $\tilde{b}_{\rm{A}}$
and $\tilde{b}_{\rm{V}}$ vanish at the tree level and are of order $10^{-2}$ at 1--loop level \cite{paper1}.
We have hence varied their value in the analysis in the range $-0.2 \div 0.2$ for $\tilde{b}_{\rm{A}}$
and $-0.1 \div 0.1$ for $\tilde{b}_{\rm{V}}$, without observing any statistically
significant variation in our results.

\begin{figure}[htb]
\vspace{4.0cm}
\includegraphics{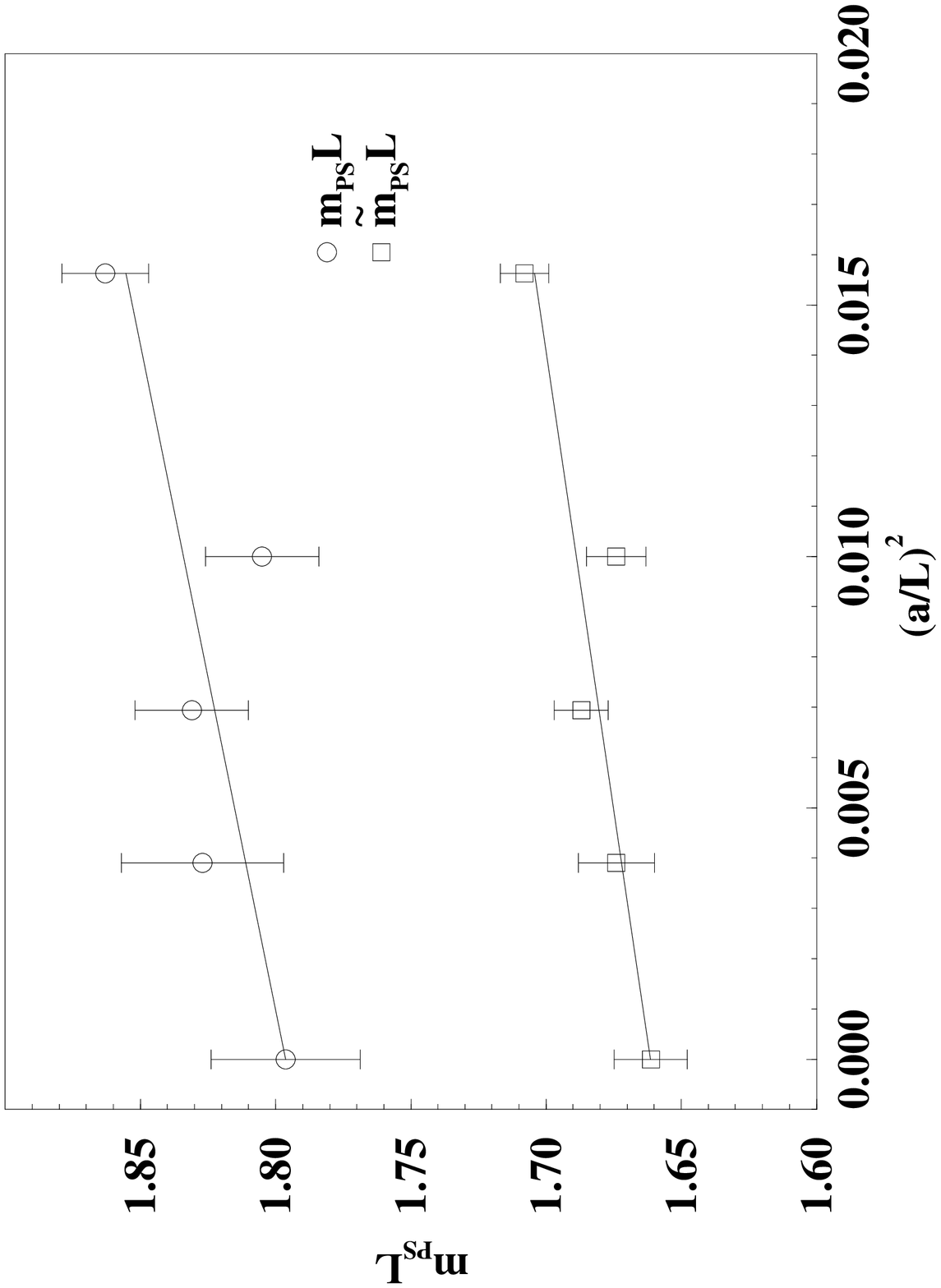}
\null\vskip 0.3cm
\caption{{\em Scaling of $m_{\rm PS}L$ and $\widetilde{m}_{\rm PS}L$.}}
\label{fig:mps}
\vspace{-0.0cm}
\end{figure}

\begin{figure}[htb]
\vspace{4.0cm}
\includegraphics{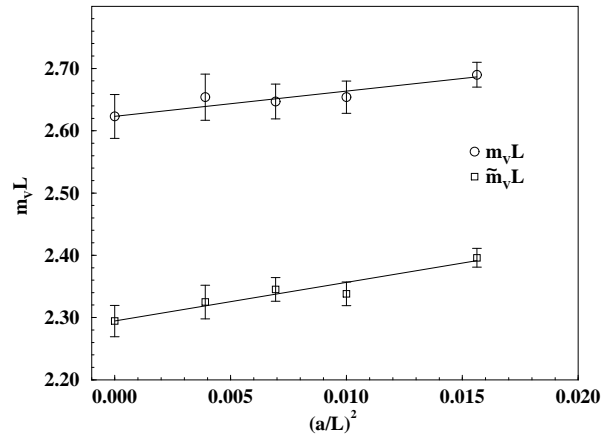}
\null\vskip 0.3cm
\caption{{\em Scaling of  $m_{\rm V}L$ and $\widetilde{m}_{\rm V}L$.}}
\label{fig:mV}
\vspace{-0.5cm}
\end{figure}

\begin{figure}[htb]
\vspace{4.0cm}
\includegraphics{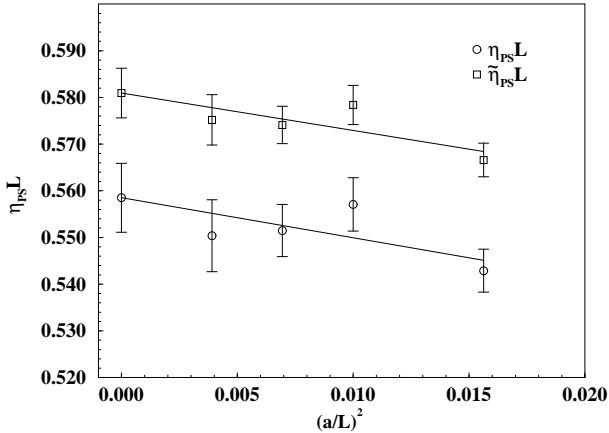}
\null\vskip 0.3cm
\caption{{\em Scaling of $\eta_{\rm PS}L$ and $\widetilde{\eta}_{\rm PS}L$.}}
\label{fig:etaps}
\end{figure}

\begin{figure}[htb]
\vspace{4.0cm}
\includegraphics{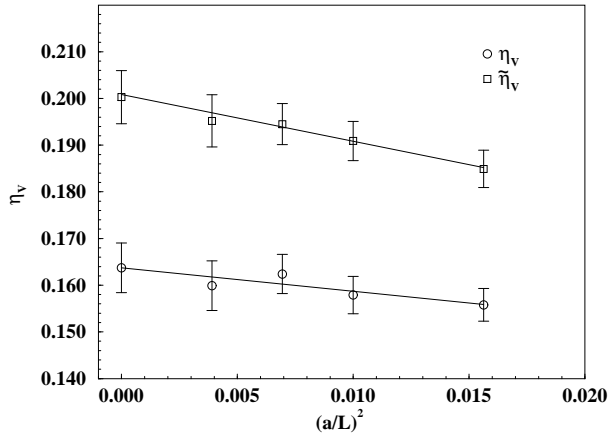}
\null\vskip 0.3cm
\caption{{\em Scaling of $\eta_{\rm V}$ and $\widetilde{\eta}_{\rm V}$.}}
\label{fig:etaV}
\vspace{-0.5cm}
\end{figure}

Our simulation points are listed in table \ref{tab:sim1}: the four values of $\beta$ correspond to
$L/a = 8, 10, 12, 16$. The quoted error on $Lm_\rmR$ is purely statistical, due to the negligible uncertainty on
$Z_{\rm A}/Z_{\rm P}$. As $Z_{\rm P}\equiv Z_{\rm P}(L_0)$
is known with a relative uncertainty of about $0.5\%$, the condition (\ref{mu_ren}) 
is implemented with this precision by adjusting $\muq$.
We correct for small mismatches with the condition eq.~(\ref{m_ren}) using numerical
estimates of the $Lm_{\rmR}$--dependence of our observables obtained via some extra simulations at $\beta=6$. 
We finally extrapolate our data to the continuum limit assuming convergence with a rate $\propto a^2$.
Our fits are shown in figures~\ref{fig:mps}-~\ref{fig:etaV}, where only statistical errors are displayed. 
The results are compatible with $O(a^2)$ deviations from the continuum limit.

Because of the finite value of $T \simeq 1.5$~fm, the observables~(\ref{obs1})--(\ref{obs8}) still
depend on the SF boundary action and fields. This induces residual O($a$) effects due to the imperfect
knowledge of $c_{\rm t}$ and $\tilde{c}_{\rm t}$ and a further 
coefficient\footnote{Due to $L\mu_{\rmR} = 0.153 \ll 1$, this coefficient was set 
to its  tree level value and never varied.} that is associated with O($a\muq$) effects \cite{paper1}.
Following~\cite{Jo}, we have performed a few extra simulations at $\beta=6$ and $\beta=6.26$ 
with values of $c_{\rm t}-1$ and  $\tilde{c}_{\rm t}-1$ that are about 2 and 10 times, respectively, 
larger than the 1--loop values. The discrepancies with the previous results
are at most about two standard deviations (at $\beta=6$) and can be considered 
negligible for the purposes of this study.

\small
\begin{table}[htb]
\vspace{-0.5cm}
\begin{tabular}{cccc}
\hline
  $m_{\rm PS}L$ &  $m_{\rm V}L$   & $\eta_{\rm PS}L$   & $\eta_{\rm V}$   \\
\hline
 1.80(3)  &2.62(4)  & 0.559(7)[1] &  0.164(5)[1] \\
   3.7\%    & 2.8\%     & 2.6\%          &  4.8\%    \\
\hline
  $\widetilde{m}_{\rm PS}L$ &  $\widetilde{m}_{\rm V}L$   & $\widetilde{\eta}_{\rm PS}L$   & $\widetilde{\eta}_{\rm V}$   \\
\hline
 1.66(1)  &2.29(3)  & 0.581(5)[1] &  0.200(6)[1] \\
   2.7\%     &  4.4\%    & 2.5\%          &  7.7\%    \\
\hline
\end{tabular}
\caption{{\em Continuum limits and deviations from $\beta=6$. 
Errors due to $Z_{\rm X}$--uncertainties in square brackets.}}
\label{tab:dev1}
\vspace{-0.7cm}
\end{table}
\normalsize

In table~\ref{tab:dev1} we compare the estimated continuum limit value of our observables with 
the value at $\beta=6$. The deviations from the continuum limit appear to be fairly small and
not larger than the ones found in an analogous study at $\muq=0$~\cite{Jo}. 


\section{Conclusions}

In the parameter region specified by $\beta \geq 6$, $L\mu_{\rm R}=0.153 \gg Lm_{\rmR}=0.020$ and $T=2L\simeq 1.5$~fm
the O($a$) improvement programme of tmQCD has been successfully implemented and tested for a few meson observables.
The renormalization of the twisted mass parameter is easy in practice, as it can be traced back to 
the renormalization of the non-singlet pseudoscalar density in the massless theory.
 
This work is part of the ALPHA collaboration research programme. 
M.~Della~Morte and R.~Frezzotti thank MURST for support.
S.~Sint acknowledges support by the European Commission under grant No. FMBICT972442.
We thank DESY and INFN for allocating computer time on APE100 machines to this project.


\begin{thebibliography}{9}
\bibitem{FGSW} R. Frezzotti, P. A. Grassi, S. Sint and P. Weisz, Nucl. Phys. B (Proc. Suppl.) 83-84 (2000) 941-946. 
\bibitem{O(a)} M. L\"uscher, S. Sint, R. Sommer and P. Weisz, Nucl. Phys. B478 (1996) 365.
\bibitem{paper1}  R. Frezzotti, P. A. Grassi, S. Sint and P. Weisz, to appear.
\bibitem{Sym} K. Symanzik, Nucl. Phys. B266 (1983) 187 and 205.
\bibitem{NP_O(a)} M. L\"uscher et al.,  Nucl. Phys. B491 (1997) 323.
\bibitem{O(a)2}  M. L\"uscher, S. Sint, R. Sommer and H. Wittig, Nucl. Phys. B491 (1997) 344.
\bibitem{Jo} J. Heitger, Nucl. Phys. B557 (1999) 309.
\bibitem{SchSte} S. Sint, Nucl. Phys. B421 (1994) 135.
\bibitem{O(a)pert}  S. Sint and P. Weisz, Nucl. Phys. B502 (1997) 251. 
\bibitem{Lu77} M. L\"uscher, Commun. Math. Phys. 54, 283 (1977).
\bibitem{SFYM} M. L\"uscher, R. Narayanan, P. Weisz and U. Wolff, Nucl. Phys. B384 (1992) 168.
\bibitem{r0scale} M. Guagnelli, R. Sommer and H. Wittig, Nucl. Phys. B535 (1998) 398.
\bibitem{ZP} S. Capitani, M. L\"uscher, R. Sommer and H. Wittig Nucl. Phys. B544 (1999) 669.
\end{thebibliography}
\end{document}